\documentclass[12pt]{article}

\def\bref#1{\noindent \parshape=2 0pt 440pt 8pt 432pt #1}

\begin{document}

\begin{center}

\Large {\bf ELECTRON INERTIAL EFFECTS ON RAPID ENERGY REDISTRIBUTION AT 
MAGNETIC X-POINTS} 

\normalsize

\vskip 1.0cm

\Large {\bf Short Title: Electron inertial effects at X-points} 

\vskip 1.0cm

K. G. McCLEMENTS

UKAEA Culham Division, Culham Science Centre, Abingdon,
Oxfordshire, OX14 3DB, UK; k.g.mcclements@ukaea.org.uk

\vskip 1.0cm

A. THYAGARAJA

UKAEA Culham Division, Culham Science Centre, Abingdon,
Oxfordshire, OX14 3DB, UK; a.thyagaraja@ukaea.org.uk

\vskip 1.0cm

N. BEN AYED

University of Bristol, H.H.Wills Physics Laboratory, Royal Fort, 
Tyndall Avenue, Bristol, BS8 1TL, UK; nb1392@bris.ac.uk 

\vskip 1.0cm

AND

L. FLETCHER

University of Glasgow, Department of Physics and Astronomy, Glasgow, G12 8QQ,
UK; lyndsay@astro.gla.ac.uk


\vfill\eject

\end{center}

\begin{abstract}
The evolution of non-potential perturbations to a current-free magnetic 
X-point configuration is studied, taking into account electron inertial 
effects as well as resistivity. Electron inertia is shown to have a 
negligible effect on the evolution of the system whenever the collisionless
skin depth is less than the resistive scale length. Non-potential magnetic field 
energy in this resistive MHD limit initially reaches equipartition with flow energy, 
in accordance with ideal MHD, and is then dissipated extremely rapidly, on an 
Alfv\'enic timescale that is essentially independent of Lundquist number. In agreement
with resistive MHD results obtained by previous authors, the magnetic field energy and 
kinetic energy are then observed to decay on a longer timescale and exhibit oscillatory 
behavior, reflecting the existence of discrete normal modes with finite real frequency.  
When the collisionless skin depth exceeds the resistive scale 
length, the system again evolves initially according to ideal MHD. At the end of this
ideal phase, the field energy decays typically on an Alfv\'enic timescale, 
while the kinetic energy (which is equally partitioned between ions and 
electrons in this case) is dissipated on the electron collision 
timescale. The oscillatory decay in the energy observed in the resistive case 
is absent, but short wavelength structures appear in the field and velocity 
profiles, suggesting the possibility of particle acceleration in 
oppositely-directed current channels. The model
provides a possible framework for interpreting observations of energy release 
and particle acceleration on timescales down to less than a second in the 
impulsive phase of solar flares. 
\end{abstract}

\section{Introduction}

Renewed interest in the long-standing problem of energy release in solar flares
has been kindled by the launch of the Ramaty High Energy Solar Spectroscopic 
Imager (RHESSI) spacecraft (Lin et al. 2002) and, very recently, by 
observations of exceptionally large flares (Bowler 2003). Resistive 
magnetohydrodynamics (MHD) provides the framework for most theoretical studies 
of this problem (for a recent review, see Priest \& Forbes 2000). Craig \& 
McClymont (1991, 1993) and Craig \& Watson (1992), for example, have invoked 
the resistive MHD relaxation of a two-dimensional magnetic X-point as a 
paradigm for 
energy release in flares. The magnetic equilibrium configuration in this 
analysis was potential (and thus stable), and the system was assumed to have a 
circular boundary centered on the X-point. Craig \& McClymont (1991, 1993) 
investigated the stable mode spectrum of such a configuration. A countably 
infinite set of discrete normal modes was identified, with azimuthally 
symmetric modes corresponding to topological reconnection. Both the real 
frequency and the damping rate were found to increase with the number of 
radial nodes. It was inferred that the spectrum after a sufficiently long time 
would be dominated by the mode with the longest radial wavelength, i.e. the 
lowest damping rate, regardless of the initial configuration. This damping 
rate was shown analytically to be given approximately by 
$$ \gamma_{\rm disc} \simeq {\pi^2c_{A0}\over 2R_0({\rm ln}S)^2}, \eqno (1) $$
where $S$ is the Lundquist number at the radial boundary of the system
$R=R_0$ and $c_{A0}$ is the Alfv\'en speed at that boundary, computed using
the equilibrium magnetic field. The
logarithmic dependence of the dissipation rate on $S$ is reminiscent of that
predicted by the Petschek ``fast'' reconnection model (Petschek 1964), 
although it should be noted that Petschek used a steady state approach while 
the treatment of Craig \& McClymont was time-dependent. The latter authors 
noted that, for typical solar parameters, equation (1) yields an energy 
release timescale of the order of several minutes to an hour, This is 
sufficiently rapid to explain thermal energy release in the gradual phase of 
a flare, but too slow to account for the impulsive phase: as noted by 
Tandberg-Hanssen and Emslie (1989), the thick target interpretation of flare 
hard X-ray bursts requires a field of the order of 0.02T to be completely 
annihilated in a volume of around 10$^{21}$m$^3$ every second.

Recognizing that resistive MHD may not provide a full description, several 
authors have investigated the possible role of non-MHD effects in flare energy 
release. Craig \& Watson (2003), for example, have identified exact 
reconnection solutions that include Hall current and electron inertial effects.
Such effects have been studied extensively by researchers in the 
magnetospheric community: Birn et al. (2001) have summarised the results of a 
recent project to compare computational models of reconnection in Harris-type 
geometry ranging from resistive MHD to particle-in-cell. In a recent paper 
McClements \& Thyagaraja (2004) generalized the spectral analysis of Craig \& 
McClymont (1991, 1993) to include electron inertia. A key motivation for doing 
so in the laboratory context is that reconnection events in magnetic fusion 
experiments, such as those responsible for ``sawtooth'' oscillations in 
tokamaks, cannot be explained using MHD alone. As noted by Wesson (1991), 
the observed relaxation time in a sawtooth crash can be much shorter 
than the timescale implied by purely resistive models. 
McClements \& Thyagaraja used the following form of Ohm's law:
$$ {\bf E}+{\bf v}\times{\bf B}=\eta{\bf j}+{m_e\over ne^2}{\partial{\bf j}
\over \partial t}, \eqno (2)$$
where $\eta$ is resistivity, $n$ is particle density, and $m_e$, $e$ denote
the electron mass and charge. McClements \& Thyagaraja demonstrated that the
magnetic X-point eigenmode spectrum in this case has a 
continuous component in addition to the discrete spectrum studied by Craig \&
McClymont. This continuum replaces the ideal MHD Alfv\'en continuum which 
is present when both resistivity and electron inertia are excluded.  
All of the finite frequency continuum modes have the same intrinsic damping 
rate, namely
$$ \gamma_{\rm cont} = {c_{A0}\over 2R_0S\delta_e^2}, \eqno (3) $$
where $\delta_e \equiv c/(\omega_{pe}R_0)$, $c$ being the speed of light and 
$\omega_{pe}$ the electron plasma frequency, is the collisionless skin depth 
normalized to the system size. Because $\delta_e$ is typically very small, the 
continuum mode damping 
rate given by equation (3) can be high in absolute terms. In fact, if the 
resistivity is determined by electron-ion collisions, $\gamma_{\rm cont}$ 
is essentially equal to the electron collision frequency 
$$ \nu_e \simeq 30{n_{15}\over T_6^{3/2}}\,{\rm s}^{-1}, \eqno (4) $$        
where $n_{15}$ is particle density in units of $10^{15}$m$^{-3}$ and $T_6$ is 
electron temperature in units of $10^6$K. Thus, for typical solar coronal 
parameters and Spitzer resistivity, the energy dissipation timescale 
implied by equation (3) is a fraction of a second. Statistically significant
fluctuations have been observed in flare hard X-ray emission on such 
timescales (e.g. Kiplinger et al. 1983): this suggests that non-MHD effects
may be significant in the context of flare energy release. However, in 
order to compute dissipation timescales reliably, it is appropriate to solve
an initial value problem rather than using an eigenvalue approach.
Craig \& Watson (1992) performed such an analysis using a resistive MHD model: 
in this paper we adopt a similar approach, but with the electron 
inertial term in Ohm's law taken into account. Specifically, we solve a 
set of linearized equations describing a perturbed magnetic X-point using 
parameters that range from the resistive MHD limit considered by Craig \&
Watson (1992) to the collisionless limit. Our approach differs from the 
two-fluid analyses carried out for example by Biskamp, Schwarz, \& Drake 
(1997) and Ramos, Porcelli \& Ver\'astegui (2002) in that we focus on a 
single, dominant non-MHD effect, namely electron inertia. Compared to 
resistive MHD, the problem is then characterised by only one additional 
parameter: the collisionless skin depth. 
Using this model, we address the key issue of how rapidly the energy in a 
non-potential magnetic field perturbation is dissipated or converted to 
kinetic energy, making an exact comparison with the resistive MHD scenario 
investigated by Craig \& Watson (1992).      
          
This paper is structured as follows. After formulating the linearized initial
value problem in \S 2, we proceed to solve it analytically and numerically  
for various parameter regimes in \S 3. The possible relevance of these 
solutions to energy release in solar flares is discussed in \S 4.    

\section{Formulation of Initial Value Problem}

We consider inviscid incompressible perturbations of a two-dimensional
current-free magnetic X-point in the low plasma beta limit. With Ohm's law 
given by equation (2), the induction and momentum equations can be written in 
the form (McClements \& Thyagaraja 2004)
$$ {\partial\over\partial t}\left(\psi-{c^2\over\omega_{pe}^2}\nabla^2\psi
\right)+({\bf v}\cdot\nabla)\psi={\eta\over\mu_0}\nabla^2\psi, \eqno (5) $$
$$ {\partial {\bf v}\over \partial t}+({\bf v}\cdot\nabla){\bf v} = 
-{1\over \mu_0\rho}(\nabla^2\psi)\nabla\psi, \eqno (6) $$
where $\psi$ is a magnetic flux function whose curl yields the  
magnetic field, $\rho$ and {\bf v} denote fluid density and velocity, and $\mu_0$ is vacuum 
permeability. Both the magnetic field and ${\bf v}$ are assumed to lie in the 
($x,y$) plane of the X-point. We linearize these equations by neglecting terms of second order 
in {\bf v} and putting $\psi=\psi_E+\tilde{\psi}$ where 
$$\psi_E = {B_0\over 2R_0}(y^2-x^2)=-{B_0R^2\over 2R_0}\cos 2\theta, \eqno (7) 
$$
$\tilde{\psi}$ is assumed to be azimuthally symmetric, 
and $\vert\nabla\tilde{\psi}\vert$ is assumed to be much smaller than
$\vert\nabla\psi_E\vert$. In equation (7) $B_0$ is the unperturbed magnetic 
field in the ($x,y$) plane at radius $R=R_0$ and $\theta$ denotes azimuthal 
angle. When linearized, equations (5) and (6) become
$$ {\partial\over\partial t}\left(\tilde{\psi}-{c^2\over\omega_{pe}^2}
\nabla^2\tilde{\psi}\right)+({\bf v}\cdot\nabla)\psi_E={\eta\over\mu_0}
\nabla^2\tilde{\psi}, \eqno (8) $$
$$ {\partial {\bf v}\over \partial t} = -{1\over \mu_0\rho_0}(\nabla^2
\tilde{\psi)}\nabla\psi_E, \eqno (9) $$
where $\rho_0$ denotes the unperturbed density (assumed to be uniform).
These equations can be combined to give a single equation for $\tilde{\psi}$ 
in which $\theta$ does not appear except in the Laplacian operator (Craig \& 
McClymont 1991; McClements \& Thyagaraja 2004). In the linear approximation, 
the variation of $\tilde{\psi}$ with $\theta$ can thus be separated  
from the variation with $R$ and $t$. In this paper we concentrate 
on the azimuthally symmetric case in which $\tilde{\psi}$ is independent of 
$\theta$. 

It is clear from equations (7) and (9) that the components of 
${\bf v}$ are $\theta$-dependent even when $\tilde{\psi}$ is azimuthally
symmetric. Formally integrating equation (9) with respect to time and writing 
the result explicitly in terms of $(R,\theta,z)$ components, we obtain
$$ {\bf v} = {B_0R\over \mu_0\rho_0R_0}(\cos 2\theta,-\sin 2\theta,0)\int
\nabla^2\tilde{\psi}dt \equiv v(\cos 2\theta,-\sin 2\theta,0). \eqno (10) $$  
The function $v(R,t)$, defined in this way, satisfies the scalar momentum
equation
$$ {\partial v\over \partial t} = {B_0R\over \mu_0\rho_0R_0}\nabla^2
\tilde{\psi}, \eqno (11) $$ 
and the radial and azimuthal velocity components are given by 
$$ v_R = v\cos 2\theta, \eqno (12) $$
$$ v_{\theta} = -v\sin 2\theta. \eqno (13) $$
It should be noted that $v$ in these expressions can be either positive or 
negative. The induction equation then becomes
$$ {\partial\over\partial t}\left(\tilde{\psi}-{c^2\over\omega_{pe}^2}
\nabla^2\tilde{\psi}\right)-{vB_0R\over R_0}={\eta\over\mu_0}
\nabla^2\tilde{\psi}. \eqno (14) $$
Following Craig \& McClymont (1991), we make the system dimensionless by 
normalising $R$ to $R_0$, $v$ to $c_{A0} \equiv B_0/(\mu_0\rho_0)^{1/2}$, $t$ to 
$R_0/c_{A0}$ and $\tilde{\psi}$ to $B_0R_0$. Dropping the 
tilde on $\tilde{\psi}$, we thus obtain
$$ {\partial \over \partial t}\left[\psi-{\delta_e^2\over r}{\partial\over
\partial r}\left(r{\partial\psi\over \partial r}\right)\right]=vr+
{1\over Sr}{\partial\over\partial r}\left(r{\partial\psi\over \partial r}
\right), \eqno (15) $$
$$ {\partial v\over \partial t} = {\partial\over\partial r}\left(r{\partial
\psi\over \partial r}\right), \eqno (16) $$
where $S \equiv \mu_0R_0c_{A0}/\eta$. 
We proceed to solve these equations subject to initial conditions on $\partial
\psi/\partial r$ and $v$, and boundary conditions on $\partial\psi/\partial r$.
Regularity requires that the perturbation to the azimuthal magnetic field
$-\partial\psi/\partial r$ vanish at $r=0$; this is consistent with the linear 
approximation since the equilibrium field in the $(x,y)$ plane has a null at 
the X-point. We also set $\partial\psi/\partial r=0$ at
$r=1$, for reasons that will be explained in the next section. Following
Craig \& Watson (1992), we consider both spatially extended initial field 
perturbations and localized ones. A simple obvious choice of initial condition 
for $v$, adopted throughout this paper, is $v=0$. 

\section{Solution of Initial Value Problem}

\subsection{Energy Evolution}

Our principal objective is to compute the evolution of field and
kinetic energy associated with a perturbed X-point. In the general case,
with finite collisionless skin depth $\delta_e$, the total energy ${\cal E}$ has
three components: non-potential field energy ${\cal E}_{\rm f}$, ion kinetic energy
${\cal E}_{\rm ki}$ and electron kinetic energy ${\cal E}_{\rm ke}$.       
In terms of the dimensionless variables introduced in the previous 
section, these energy components can be represented by the integrals
$$ {\cal E}_{\rm f} = {1\over 2}\int_0^1\left({\partial\psi\over
\partial r}\right)^2rdr, \eqno (17) $$
$$ {\cal E}_{\rm ki} = {1\over 2}\int_0^1v^2rdr, \eqno (18) $$
$$ {\cal E}_{\rm ke} = {1\over 2}\int_0^1\delta_e^2(\nabla^2\psi)^2rdr. 
\eqno (19) $$
The identification of the integral in equation (17) as the non-potential field 
energy follows from the fact that the magnetic field perturbation is $ -\partial
\psi/\partial r$. Locally, there is a contribution to the field energy density 
proportional to $(\partial\psi_E/\partial r)(\partial\psi/\partial r)$, but this gives a
zero contribution to the total energy when integrated over $\theta$ since
$\partial\psi_E/\partial r$ is proportional to cos$2\theta$ whereas
$\partial\psi/\partial r$ is azimuthally symmetric. Ions make the dominant 
contribution to the bulk fluid velocity {\bf v}, and therefore it is appropriate to regard the 
integral in equation (18) as a measure of ion kinetic energy. On the other hand, 
most of the current is carried by electrons, and the current density 
(which is oriented in the $z$-direction) is proportional to $-\nabla^2\psi$:
it follows from this that the integral in equation (19) represents electron kinetic 
energy. 

In the limit $S \to \infty$ the total energy
$$ {\cal E} = {\cal E}_{\rm f}+{\cal E}_{\rm ki}+{\cal E}_{\rm ke}=
{1\over 2}\int_0^1\left[\left({\partial\psi\over\partial r}\right)^2+v^2+
\delta_e^2(\nabla^2\psi)^2\right]rdr, \eqno (20) $$
is conserved. To establish this result, we evaluate the time derivative of 
${\cal E}$, using equations (15) and (16) to eliminate $\partial\psi/
\partial t$ and $\partial v/\partial t$, obtaining 
$$ {d{\cal E}\over dt} = -{1\over S}\int_0^1\left(\nabla^2\psi\right)^2rdr, 
\eqno (21) $$
where we have invoked the boundary conditions $\partial\psi/\partial r=0$ at 
$r=0$ and $r=1$. The right hand side of equation (21) is simply the rate of
Ohmic heating. By setting $\partial\psi/\partial r=0$ at $r=1$, we ensure 
that the Poynting flux is locally zero at all points on the boundary. We are 
not setting $v=0$ at $r=1$, and therefore we are allowing the possibility of a 
local mass flow through the boundary. However, since the radial component of the velocity 
vector, like the equilibrium field, has a $\cos 2\theta$ dependence (eq. [12]),
the integrated mass flux through the boundary is always zero. Thus, after $t=0$ there is no net
flow of energy into or out of the system, and the energy can only decrease, 
through Ohmic dissipation, with the rate of decrease determined purely by the 
internal dynamics of the system. 

It is clear from equation (21) that the system is conservative in the limit $S \to 
\infty$, even when $\delta_e$ is finite: electron inertial effects are essentially 
reactive and thus do not give rise to dissipation. However, we will demonstrate 
that the field energy ${\cal E}_{\rm f}$ can change rapidly when $\delta_e$ is 
finite and $S \to \infty$. The conservation of ${\cal E}$ in this limit provides a 
critical test of the numerical schemes used for solving the full equations. In 
Appendix A we obtain a relationship between field energy and electron 
kinetic energy that will prove to be useful for the interpretation of finite 
$\delta_e$ solutions presented in \S\S 3.4. 

\subsection{Ideal MHD Solutions}

As first noted by Bulanov \& Syrovatskii (1981), a simple analytical solution 
of the linearized equations exists in the ideal MHD limit ($S \to \infty$, 
$\delta_e \to 0$). This solution provides an additional test of the numerical
schemes, and moreover is key to understanding certain effects 
observed in the non-ideal case.     

Following Craig \& Watson (1992), we evaluate the ideal MHD solution explicitly 
for two alternative initial conditions.
When $\delta_e=1/S=0$ equations (15) and (16) can be combined to give 
$$ {\partial^2\psi\over\partial t^2} = r{\partial\over\partial r}\left(r
{\partial\psi\over\partial r}\right). \eqno (22) $$
With the substitution $u=\ln r$, equation (22) reduces to the one dimensional 
wave equation
$$ {\partial^2\psi\over\partial t^2} = {\partial^2\psi\over\partial u^2},
\eqno (23) $$
with general solution
$$ \psi = f(u+t)+g(u-t)=f(\ln r+t)+g(\ln r-t), \eqno (24) $$
the functions $f$ and $g$ being arbitrary. Employing the boundary conditions 
$\partial\psi/\partial r=0$ at $r=0$ and $r=1$ and initial conditions
$$ {\partial\psi\over\partial r} = \sin(\pi r), \eqno (25) $$
it is straightforward to show that $f$ and $g$ are such that the full solution
is given by
$$ \psi = -{1\over 2\pi}\left\{\cos(\pi re^t)+\cos(\pi re^{-t})\right\}, \; t <
-\ln r \eqno (26{\rm a})$$
$$ \psi = -{1\over 2\pi}\left\{\cos\left({\pi\over r}e^{-t}\right)+\cos(\pi 
re^{-t})\right\}, \; t > -\ln r \eqno (26{\rm b})$$
Putting $\delta_e=1/S=0$ in equation (15), we obtain
$$ v = {1\over 2}\left\{e^t\sin(\pi re^t)-e^{-t}\sin(\pi re^{-t})\right\}, 
\; t < -\ln r \eqno (27{\rm a})$$
$$ v = -{1\over 2}\left\{{e^{-t}\over r^2}\sin\left({\pi\over r}e^{-t}\right)+
e^{-t}\sin(\pi re^{-t})\right\}, \; t > -\ln r \eqno (27{\rm b})$$ 
The solutions for $t > -\ln r$ represent smooth continuations of those for 
$t < -\ln r$: there is no discontinuity at this point. Using these 
expressions, we can compute the time evolution of the ion kinetic and field 
energy. As shown in the left plot of Figure 1, the two 
components of the energy undergo a single oscillation before reaching 
equipartition in about three Alfv\'en times. The field and velocity profiles 
continue to evolve after this time, however. As shown in the right hand frame 
of Figure 1, the field energy and kinetic energy become increasingly 
concentrated in the vicinity of the X-point. Eventually, the field and 
velocity gradients become sufficiently large that the neglect of resistive and 
electron inertial terms in equations (15) and (16) is no longer justified and 
the ideal MHD solution becomes invalid. The right hand frame of Figure 1 
exemplifies the well-known tendency of magnetic X-points to focus 
and accrete electromagnetic and kinetic energy (see e.g. Craig \& Watson
1992). 

The trajectory of the inward-propagating wave described by equation (24)
is given by $dr/dt = -r$: this arises from the Alfv\'enic character of the wave and the
fact that the magnitude of the equilibrium magnetic field is proportional to 
$r$ (cf. eq. [7]). McLaughlin \& Hood (2004) have recently demonstrated 
an important consequence of this, namely that a disturbance initially 
consisting of a plane wave is refracted as it approaches the X-point in such a way 
that it becomes more azimuthally symmetric: the region of the wave front closest to the 
null propagates more slowly than neighboring regions. For this
reason, it is particularly appropriate to consider the evolution of 
azimuthally symmetric perturbations.                

It is also instructive to compute the ideal MHD solution for the case of a 
more localized initial flux perturbation of the form
$$ \psi = -{\Delta u^2\over 2}\exp\left[-\left({\ln r\over 
\Delta u}\right)^2\right], \eqno (28) $$ 
where $\Delta u$ is a constant characterizing the width of the disturbance 
in $\ln r$ space. Craig \& Watson (1992) considered initial perturbations
similar to this, but with a factor of $\ln r$ outside the exponential (so that
$\psi$ rather than $\partial\psi/\partial r$ was set equal to zero at $r=1$) 
and $\Delta u = 1$. Profiles of this type are only strongly localized if 
$\Delta u \ll 1$. Determining $\psi(r,t)$ for this case 
from equation (24), and using equation (15) with $\delta_e=1/S=0$ to compute 
$v$, we obtain
$$ {\partial\psi\over\partial r}={1\over 2r}\left\{(\ln r +t)\exp\left[
-\left({\ln r+t\over \Delta u}\right)^2\right]+(\ln r -t)\exp\left[
-\left({\ln r-t\over \Delta u}\right)^2\right]\right\}, \eqno (29{\rm a}) $$
$$ v={1\over 2r}\left\{(\ln r +t)\exp\left[
-\left({\ln r+t\over \Delta u}\right)^2\right]-(\ln r -t)\exp\left[
-\left({\ln r-t\over \Delta u}\right)^2\right]\right\}. \eqno (29{\rm b}) $$
If $\Delta u \ll 1$, $\partial\psi/\partial r$ and $v$ are generally
negligible except for $\ln r \sim \pm t$. Since we are only considering 
$t > 0$ and $r < 1$, i.e. $\ln r < 0$, the dominant terms in equation (29) are
those multiplied by $\ln r +t$, representing an inward propagating wave. It is
immediately apparent in this case that $v \simeq \partial\psi/\partial r$, i.e.
energy is equally partitioned between the non-potential component of the 
magnetic field and the bulk flow, as in the case of the global disturbance 
represented by equation (25).    

The energy components corresponding to this solution are plotted versus 
time for $\Delta u=0.1$ in the left hand frame of Figure 2. It is apparent 
that equipartition occurs much more rapidly than in the case of Figure 1. The 
right hand frame of Figure 2, showing the evolution of $\partial\psi/\partial 
r$, indicates as before rapid focusing of the field energy. Steepening of the 
profiles is noticeably more rapid than in the case of Fig. 1: we would 
accordingly expect non-ideal effects to occur at an earlier stage of the 
system evolution. The two solutions considered above illustrate the fact that 
energy equipartition is a natural asymptotic state of the ideal MHD system
(cf. Craig \& Watson 1992). 

\subsection{Resistive MHD Solutions}

In this subsection we present results obtained from numerical solutions of 
equations (15) and (16) for the resistive MHD case, $\delta_e=0$. This 
scenario, with somewhat different initial conditions, was investigated 
extensively by Craig \& Watson (1992): we consider it here in order to provide 
an exact benchmark for the finite $\delta_e$ cases considered in \S\S 3.4, and 
also to highlight a phase in the late evolution of ${\cal E}$ not discussed by 
those authors. Independently-written codes were used to solve equations (15) 
and (16) for $\delta_e=0$ and for finite $\delta_e$ (\S\S 3.4).
The numerical methods used in these codes are described briefly in Appendix B.   

Our initial conditions are those defined by equations (25) and (28), together
with $v=0$, as before. In the case of the more localized field perturbation
(eq. [28]), we take $\Delta u = 0.1$. For the two types of initial field 
perturbation, we have determined the evolution of the system for a wide range 
of Lundquist numbers $S$, from $10$ to $10^8$; Craig \& Watson (1992) 
concentrated their investigation on the case of $S=10^8$. Our results, shown 
in Figures 3 and 4, concur at high $S$ with those obtained by Craig \& Watson. 
For $S > 10^3$ there is an early phase in which the system evolution is 
well-described by ideal MHD. In Figure 3(h), for example ($S=10^8$), the time 
evolution of the energy components up to $t \simeq 7$ is identical to that 
computed using the corresponding ideal solution (Figure 1).
The ideal phase becomes progressively longer as $S$ is increased: 
for very small $S$, on 
the other hand, there is no ideal phase and the energy decays immediately. 
Although the onset time of the drop in energy increases with $S$, the timescale
of the decay itself is essentially independent of this parameter, and is 
comparable to or less than one Alfv\'en time. The factor by which the energy 
falls at this point appears to depend critically on
the initial field perturbation: this was also found by Craig \& Watson. 
In the case of the more localized perturbation (Fig. 4), the energy is
reduced in this phase by a factor of about $e^{10} \simeq 2 \times 10^4$.   

The rapid drop in energy is followed by a period of slower decay in which the 
field energy and kinetic energy exhibit oscillatory behavior. As noted by
Craig \& Watson (1992), these oscillations arise from the 
discrete spectrum of weakly damped eigenmodes studied by Craig \& McClymont 
(1991).  After a sufficiently long period, the oscillations disappear. This 
can be seen in Figures 3 and 4 for $S \le 10^3$, and in Figure 5 for $S=10^4$, 
$10^5$. The oscillations persist for a time that increases as $(\ln S)^2$: 
this is consistent with the $S$ dependence of the least damped discrete mode 
(eq. [1]).    

After the oscillatory phase, the energy continues to decay, but on a  longer 
timescale, and the energy moreover is almost entirely kinetic 
(${\cal E}_{\rm ki} \gg {\cal E}_{\rm f}$). This late phase of non-oscillatory 
decay, which is apparent in Figures 3(a), 3(b), 4(a), 4(b) and 5, 
can be accounted for by the fact that
the discrete modes identified by Craig \& McClymont (1991) do not constitute
a complete set: as demonstrated by McClements \& Thyagaraja (2004), there is 
also a continuum of damped modes with zero real frequency. When $\delta_e$ and 
$S$ are finite, equation (22) is replaced with the fourth order equation
$$ \ddot{\psi}-{\delta_e^2\over r}{\partial\over\partial r}\left(r
{\partial\ddot{\psi}\over\partial r}\right) = r{\partial\over\partial r}\left(r
{\partial\psi\over\partial r}\right)+{1\over Sr}{\partial\over\partial r}
\left(r{\partial\dot{\psi}\over\partial r}\right). \eqno (30) $$    
For $\psi = e^{-i\omega t}f(r)$, equation (30) reduces to 
$$\omega^2 rf + \left(r^2-\omega^2\delta_e^2-i{\omega\over S}\right){d\over 
dr}\left(r{df\over dr}\right)=0. \eqno (31) $$
The coefficient of the highest order derivative in this equation vanishes
for complex mode frequencies $\omega$ satisfying
$$ \omega^2\delta_e^2+i{\omega\over S} = r^2. \eqno (32)$$
McClements \& Thyagaraja showed that this relation gives rise to two distinct 
continua: one with finite real frequency, the other with zero real frequency. 
The former, which we will discuss in the next section, only exists when 
$\delta_e \ne 0$. The zero real frequency continuum, on the other hand, exists
whether $\delta_e$ is finite or not. This can be seen by putting $\omega = 
-i\gamma$, where $\gamma$ is taken to be real. Equation (32) then yields
$$ \gamma = {1\over 2S\delta_e^2}\left[1 \pm\left(1-4r^2S^2\delta_e^2\right)^{1/2}
\right]. \eqno (33) $$
In the limit $S^2\delta_e^2 \ll 1$, the negative root in this expression
yields a damping rate 
$$ \gamma = Sr^2. \eqno (34) $$
Thus, there is a purely damped continuum mode for each point in the 
solution domain. Since this extends to $r=0$, the damping can be arbitrarily
weak.  In the resistive MHD case, the Fourier spectra of $\psi$ and $v$ will
include contributions from this continuum as well as the discrete modes
identified by Craig \& McClymont. Equation (34) indicates that non-potential
perturbations to the X-point field will in general take an infinite time to be
completely dissipated. The presence of the continuum only becomes apparent
when the discrete modes have disappeared: since the discrete mode
damping rate scales with $1/(\ln S)^2$ (eq. [1]), the presence of the 
continuum becomes manifest at progressively later times as $S$ is increased.   

As noted above, another feature of the late non-oscillatory decay phase is 
that the remaining energy in this period is almost entirely kinetic. This can
be understood qualitatively by putting $\omega = -i\gamma$ and writing equation
(31) in the form
$$ \left(rf^{\prime}\right)^{\prime} = {\gamma^2rf\over r^2-\gamma/S}, 
\eqno (35) $$
where primes denote derivatives with respect to $r$.
Assuming that $f$ tends to a finite value as $r$ approaches the singular
point $(\gamma/S)^{1/2}$, as in the case of the finite real frequency continuum
modes discussed by McClements \& Thyagaraja (2004), we infer from 
equation (35) that whereas the magnetic field perturbation $f^{\prime}$ 
diverges logarithmically in the neighborhood of $r=(\gamma/S)^{1/2}$, the 
current $(rf^{\prime})^{\prime}/r$ and hence the velocity (cf. eq. [16]) 
diverge as $1/[r-(\gamma/S)^{1/2}]$. For these continuum eigenfunctions, the 
field energy is thus square integrable whereas the kinetic energy is not. The
energy components in the late non-oscillatory phase of the resistive MHD 
simulations can, in principle, be synthesized using these singular eigenmodes.
In view of the nature of the singularities in $f^{\prime}$ and 
$(rf^{\prime})^{\prime}/r$, it is not surprising that the kinetic energy, 
although necessarily finite (as indicated by eq. [21], the total energy can 
never increase), is large compared to the field energy.  
       
\subsection{Finite Collisionless Skin Depth Solutions}

We have obtained numerical solutions of equations (15) and (16) with $\delta_e
= 0.01$ and $S$ ranging from $10^2$ to $10^6$, using both types of initial
condition discussed in the previous two subsections. Figure 6 shows the 
evolution of the field energy, ion kinetic energy and electron kinetic energy
in each case, with the left hand and right hand plots corresponding 
respectively to the spatially extended initial condition (eq. [25]) and the
strongly localized initial condition (eq. [28]). 
Comparing Figure 6 with Figures 3 and 4, it can be seen that 
electron inertia has a negligible impact on the evolution of 
${\cal E}_{\rm f}$, ${\cal E}_{\rm ki}$ and ${\cal E}$ whenever $S\delta_e^2 
< 1$: the black, red and blue curves in Figure 3(b) and the left hand frame 
of Figure 6(a), for example, which were obtained using the same initial
conditions and $S=10^2$, are indistinguishable. This is
particularly noteworthy since the results shown in Figure 6 and those discussed
in \S\S 3.3 were obtained using two independent codes (see Appendix B). 
When $S\delta_e^2 = 0.01$, ${\cal E}_{\rm ke}$ remains below the
other energy components at all times for both sets of initial conditions 
(Fig. 6[a]). When the initial field perturbation has the profile $\sin\pi r$,
the ratio ${\cal E}_{\rm ke}/{\cal E}_{\rm f}$ is well-approximated at $t=0$ 
by the equals sign in equation (A7). The ratio is somewhat higher in the case 
of the more localized initial field perturbation: this appears to be due to a 
greater contribution of eigenvalues $\lambda_n^2 > \lambda_0^2$ in equation 
(A8). Towards the end of simulations with $S\delta_e^2 = 0.01$, 
${\cal E}_{\rm ke}$ becomes comparable to the field energy, but the system
evolution is still well-approximated by resistive MHD. 

The situation is completely different when $S\delta_e^2 > 1$: this regime
is represented by Figures 6(d) and 6(e), with Figure 6(c) representing a
transitional case ($S\delta_e^2=1$). The oscillatory phase prevalent in the 
resistive MHD simulations is still present when $S\delta_e^2 = 1$ but wholly 
absent when this quantity is significantly greater than unity. After a 
few Alfv\'en times, the total energy is dominated by ions and electrons, with 
${\cal E}$ equally partitioned between the two species. The total energy
still decays, as it must according to equation (21), but at a diminishing 
rate as $S$ increases. In fact, the asymptotic decay time of the 
total energy is equal to $S\delta_e^2$. Given that ${\cal E}$ is dominated 
by electrons and ions, with ${\cal E}_{\rm ki} \simeq {\cal E}_{\rm ke}$
so that ${\cal E}_{\rm ke} \simeq {\cal E}/2$, the decay rate follows very 
simply from equation (21):
$$ {d{\cal E}\over dt} = -{1\over S\delta_e^2}\int_0^1\delta_e^2
\left(\nabla^2\psi\right)^2rdr = -{2{\cal E}_{\rm ke}\over S\delta_e^2} \simeq 
-{{\cal E}\over S\delta_e^2}. \eqno (36) $$         
McClements \& Thyagaraja (2004) demonstrated that the discrete X-point
spectrum ceases to exist when the intrinsic damping rate of the 
finite frequency continuum is less than that of the least damped discrete 
mode. As noted previously, the latter varies as $1/(\ln S)^2$ and is of order 
unity in Alfv\'en units for low $S$. Thus, when $S\delta_e^2 \gg 1$ the 
spectrum is purely continuous. The decay rate indicated by equation (36) 
is exactly twice the intrinsic damping rate of finite real frequency continuum 
eigenmodes (cf. eq. [3]). This is to be expected if the energy of the system is
assumed to lie predominately in the finite frequency continuum, since the
ion and electron kinetic energies scale as the squares of, respectively, the 
fluid velocity and the current. As noted in \S 1, the energy decay time 
$S\delta_e^2$ is simply the electron collision time if $S$ is determined by 
Spitzer resistivity. 

For these simulations there is a period in which the field energy decays
much more rapidly than this, typically on an Alfv\'en timescale: this causes
${\cal E}_{\rm f}$ to be energetically insignificant at later times. As in the 
later stages of the resistive MHD simulations, the dominance of kinetic energy 
over field energy when $S\delta_e^2 > 1$ is associated with the fact
that for continuum eigenfunctions the singularity in the field is square 
integrable while those in the current and velocity are not (McClements \& 
Thyagaraja 2004). The only difference in this case is that part of the 
continuum has finite real frequency. 

The field energy decays on the Alfv\'en timescale even in the collisionless 
limit ($S\delta_e^2 \to \infty$): we thus observe irreversible behavior, 
despite the fact that total energy is conserved in this case. As demonstrated 
by McClements \& Thyagaraja (2004), the spectrum in the limit $S\delta_e^2 \to 
\infty$ consists of a band-limited continuum. The decay of field energy in this
limit can be interpreted as continuum damping. It is exactly analogous to 
Landau damping: the Vlasov-Poisson system of equations, like the system 
represented by equations (15) and (16) in the limit $S\delta_e^2 \to \infty$, 
is conservative but exhibits irreversible behavior (see e.g. van Kampen \& 
Felderhof 1967). The analysis of inviscid plane Couette flow by Case (1960) 
shows that this type of behavior is not restricted to plasma kinetic theory
or wave-particle interactions but is a general consequence of phase mixing. 
Thyagaraja et al. (2002) have demonstrated that even the simplest 
advection-diffusion equation in one space dimension can exhibit such 
dissipationless damping. In the solar context, phase mixing of MHD waves
has been investigated by many authors as a possible coronal heating mechanism
(e.g. Heyvaerts \& Priest 1983; Voitenko \& Goosens 2000; Tsiklauri, 
Nakariakov, \& Rowlands 2003). Porcelli et al. (2002) have shown that the 
channeling of field energy into kinetic energy through phase mixing occurs in 
the collisionless evolution of magnetic islands. In the case of finite $S$ 
with $S\delta_e^2>1$, as in Figures 6(d) and 6(e), the field energy decays 
through a combination of phase mixing and the intrinsic finite frequency 
continuum mode damping given by equation (3).

Another characteristic of the $S\delta_e^2 > 1$ regime indicated by Figure 6 is
that the evolution of ${\cal E}$ is relatively insensitive to the initial
conditions: in Figures 6(d) and 6(e), with $S\delta_e^2$ equal to 10 and 
100 respectively, the left hand plots are almost identical to the right hand
plots. In the case of the more localized perturbation, the electrons and ions 
take somewhat less time to reach equipartition, but the subsequent evolution of
the energy components is essentially independent of the initial field profile. 
In the resistive MHD regime, in contrast, it is clear from Figures 3 and 4 that
the total decrement in the energy after a given time interval depends 
critically on the initial conditions, although the rate of energy loss does 
not.

The profiles of $\partial\psi/\partial r$ and $v$ also depend critically on 
whether $S\delta_e^2$ is less than or greater than unity. Figure 7 shows the 
evolving velocity profile up to $t=4$ for the case of initial $\psi \propto 
-\exp[-(\ln r)^2/\Delta u^2]$, $\delta_e=0.01$ and (a) $S=10^3$, (b) $S=10^5$. 
For $S\delta_e^2 < 1$, the profile steepening observed in the ideal phase 
(Figs. [1] and [2]) ceases due to resistive effects, and the field and 
velocity are damped without the shapes of their spatial profiles being 
radically altered. A low amplitude ripple appears in the wake of the inward 
propagating pulse, apparently due to finite $\delta_e$ (the ripple is absent 
in the resistive MHD limit), but, as indicated by Figure 8(a), the overall 
profile shape is similar to the ideal MHD profiles shown in the right hand 
frame of Figure 2. When $S \delta_e^2 > 1$, on the other hand, the field and 
velocity profiles are characterized by short wavelength features that persist 
for long times (Figs. 7[b] and 8[b]). Porcelli et al. (2002) observed a similar
filamentation process in collisionless simulations of a magnetic island. 
The appearance of progressively more nodes in the field profile means that the 
Rayleigh quotient defined by equation (A2) increases with time: this is linked 
to the rapid decay of field energy relative to electron kinetic energy noted 
above. Figures 7(b) and 8(b) indicate that the radial 
wave number spectrum is cascading into progressively shorter wavelengths. 
Examples of such direct cascades investigated previously in the plasma 
literature (e.g. by Thyagaraja, Loureiro, \& Knight 2002) show that they play 
a crucial role in the temporal evolution of field energy. 

The occurrence and persistence of small spatial scales for $S\delta_e^2 > 1$ 
is linked to the fact that the only eigenmodes in 
this limit are singular, and are thus characterized by arbitrarily short 
wavelengths: the energy in the initial perturbation is channeled into these 
short wavelengths, giving rise to the
observed profiles. McClements \& Thyagaraja (2004) showed that the 
finite frequency continuum is present even when $S\delta_e^2 < 1$, but in this
case the component of the energy in this continuum rapidly damps out 
(on the $S\delta_e^2$ timescale), and the remaining energy decays on 
a timescale determined by the discrete mode damping rate. Since the discrete
modes are non-singular, they are not characterised by short wavelengths.
Thus, in the resistive MHD regime we observe relatively smooth 
spatial profiles (Figs. 7[a] and 8[a]) but highly structured evolution of the 
total energy (Figs. 3, 4, 6[a] and 6[b]), whereas in the $S\delta_e^2 > 1$ 
regime the profiles have fine scale structure (Figs. 7[b] and 8[b]) but the 
temporal behavior of the total energy is relatively smooth (Figs. 6[d] and 
6[e]). In the latter case 
short wavelength structures also appear in the current profile and the 
longitudinal component of the electric field, $-\partial\psi/\partial t$. 
Electrons (and ions) would then be accelerated in nested cylindrical shells, 
centered on the X-line, with acceleration occurring in opposite directions in 
adjacent shells. One attractive feature of this scenario is that it suggests 
the possibility of producing large numbers of hard X-ray emitting electrons 
without the creation of large net beam currents, and hence unacceptably large
magnetic fields (see e.g. Holman 1985).           

McClements \& Thyagaraja (2004) noted that the discrete X-point spectrum 
found by Craig \& McClymont (1991) is replaced in the ideal limit with a continuum: 
the eigenmodes of this continuum, like those of the non-ideal continua discussed 
above, are singular. The cascading of field and kinetic energy into small spatial scales 
apparent in Figures 1 and 2 is associated with the singular nature of the ideal
MHD eigenmodes. When $S$ is finite but $\delta_e=0$, the finite frequency ideal MHD    
spectrum becomes discrete and the eigenmodes non-singular. However, as
discussed in \S\S 3.3, the system evolution at high $S$ is still accurately described by 
ideal MHD for several Alfv\'en periods after $t=0$: energy cascades into small scales 
in this period, despite the resolution of the continuum into discrete 
modes. Similarly, although one would expect the non-ideal continua that exist for finite $S$ and 
$\delta_e$ to be resolved into discrete spectra by, for example, finite gyro radius 
effects, such effects will not necessarily prevent  filamentation or the decay of field 
energy relative to kinetic energy. On the other hand, gyro radius effects are likely 
to impose a lower limit on the length scale of the filaments: in the simulations reported 
here with $S\delta_e^2>1$, there appears to be no such lower limit.          

\section{Summary and Application to Solar Flares}

We have studied the relaxation of non-potential perturbations to a current-free
magnetic X-point, taking into account the effects of resistivity and electron 
inertia. The latter has been shown to have a negligible effect on the 
evolution of the system whenever the collisionless skin depth is less than the 
resistive scale length. Non-potential magnetic field energy in this resistive MHD limit 
initially reaches equipartition with flow energy, in accordance with analytical results
obtained using ideal MHD, and is then dissipated extremely rapidly, on an Alfv\'enic 
timescale that is essentially independent of Lundquist number, with the energy 
being attenuated to a greater extent if the initial perturbation is highly localized. 
Following this period of rapid dissipation in the resistive case, the magnetic field energy and 
kinetic energy decay on a longer timescale and exhibit oscillatory behavior: Craig \&
Watson (1992), who also observed such behaviour in resistive MHD X-point 
simulations, noted that it arises
from the existence of discrete normal modes with finite real frequency. 
Eventually, the oscillations disappear, and the energy continues to decay on a 
yet longer timescale. This non-oscillatory decay arises from the fact that the 
discrete modes observed in the earlier evolution do not constitute a complete 
set, even in the resistive limit; the spectrum also contains a purely damped 
continuum. When the collisionless skin depth exceeds the resistive scale 
length, the system again evolves initially according to ideal MHD. At the end of this
ideal phase, the field energy decays typically on an Alfv\'enic timescale, 
while the kinetic energy (which is equally partitioned between ions and 
electrons in this case) is dissipated on the electron collision 
timescale. The oscillatory decay in the energy observed in the resistive case 
is absent, but short wavelength structures appear in the field and velocity 
profiles, suggesting the possibility of particle acceleration in 
oppositely-directed current channels.  

We now consider the possible application of these results to the problem of 
short timescale energy release in solar flares. For a system size $R_0$ equal 
to the typical length of a flaring coronal loop ($\sim 10^7$m), 
$B_0 \sim 0.1$T, $n=10^{15}$m$^{-3}$, $T \sim 3 \times 10^6$K, and Spitzer 
resistivity, the Lundquist number $S$ is of the order of $3 \times 10^{15}$
while the dimensionless collisionless skin depth is about $10^{-8}$ and hence 
$S\delta_e^2 \simeq 1$. This indicates that the 
collisionless skin depth $\delta_e$, although very small compared to the 
typical spatial extent of a solar flare, is actually comparable to the 
resistive length scale $1/S^{1/2}$ if the latter is assumed to be determined 
by Coulomb collisions. Since Spitzer resistivity is often invoked as the source
of magnetic field energy dissipation in flares, the figures quoted above 
suggest that electron inertial effects are likely to play a role in this 
process. However, in the absence of precise knowledge of magnetic field 
structures in the pre-flare corona, there is no particular reason for 
identifying $R_0$ with a macroscopic flare dimension, given that no other 
length scales appear in the current-free X-point field defined by equation 
(7). Indeed, it is unlikely that this equation would be a good approximation 
to the coronal field over length scales as large as $10^7$m. More 
realistically, it could be regarded as a model of a spatially restricted 
region, with dimensions much smaller than a typical flare size, for which
the approximation of a current-free two-dimensional X-point is appropriate.
Non-potential magnetic field energy originating from a much more extensive     
region could still be channeled into the X-point through the focusing effect
noted by Craig \& Watson (1992). A key point to note here is that $S$ is 
defined in terms of a field component which increases linearly with distance 
from a two-dimensional null (it is independent of any longitudinal field 
component). If the boundary is chosen to be arbitrarily close to the origin, 
the Lundquist number is then arbitrarily small. In this specific context, it 
is thus entirely appropriate to consider values of $S$ much smaller than the
global Lundquist numbers characteristic of the flaring solar corona, and values
of $\delta_e$ much larger than the figure of $10^{-8}$ quoted above. 

The choice of $R_0$ is not completely arbitrary, since the boundary conditions 
imposed at this radius must have some effect on the evolution of ${\cal E}$. 
However, the results presented in the previous section and by Craig \& Watson
(1992) suggest strongly that any such effect is very weak. In the resistive 
regime, our results are essentially identical to those obtained by Craig \& 
Watson despite the use of different boundary conditions. This is consistent 
with the fact that the time-evolving field and velocity profiles in both the
resistive MHD and electron inertial regimes invariably show dissipation 
occurring much closer to the X-point than the boundary at $R=R_0$. 

Regardless of the initial conditions, our results show the greater part of
the non-potential magnetic field energy either being dissipated or, 
in the case of $S\delta_e^2>1$, being converted into kinetic energy, on the 
Alfv\'en time $\tau_A = R_0/c_{A0}$. In the event of $S\delta_e^2$ being 
greater than unity, with $S$ determined by Spitzer resistivity, the kinetic 
energy is dissipated on the electron collision time $\tau_e$. Since 
$S\delta_e^2$ is simply the ratio $\tau_e/\tau_A$ (cf. eqs. [3] and [4]), the 
overall energy dissipation timescale can thus be written as max$(\tau_A,
\tau_e)$. As noted in 
\S 1, $\tau_e$ is typically less than a second in the flaring corona. The 
Alfv\'en time could also be less than a second, if the X-point field were 
sufficiently highly sheared. Again putting $n=10^{15}$m$^{-3}$, and assuming 
that the field rises to, say, $0.01$T at a radial distance from the null of 
$10^6$m, we obtain $\tau_A \simeq 0.1$s. Since it is the destruction of 
magnetic flux that gives rise to an accelerating electric field in the 
$z$-direction $E_z = -\partial\psi/\partial t$, the key
timescale from the point of view of energetic particle production is always
$\tau_A$, provided that the acceleration time is no longer than this. The 
latter timescale depends, of course, on the amplitude of the initial magnetic 
field perturbation. Since equations (15) and (16) are linear, the perturbation 
amplitude is arbitrary as far as the results shown in Figures 1-8 are 
concerned. However, we can at least check that the quantity of magnetic flux 
$\Delta\psi$ that must be destroyed to produce, say, a 30keV electron in 0.1s 
is broadly consistent with the linear approximation. The energy ${\cal E}_e$
acquired by a nonrelativistic electron in time $t$ is
$$ {\cal E}_e \sim {e^2t^2\over 2m}{\Delta\psi^2\over \tau_A^2}, \eqno (37) $$
so that
$$\Delta\psi = (2m{\cal E}_e)^{1/2}{\tau_A\over et}. \eqno (38) $$
The perturbed field is $\tilde{B}_{\varphi} \sim \Delta\psi/R_0$ and the 
unperturbed field is $B_E \sim B_0$: their ratio is 
$$ {\tilde{B}_{\varphi}\over B_E} \sim (2m{\cal E}_e)^{1/2}{\tau_A\over etR_0
B_0}. \eqno (39) $$
Using the parameter values invoked above, we obtain $\tilde{B}_{\varphi}/B_E
\sim 10^{-7}$. Even allowing for steepening of the perturbed field during the 
course of each simulation, this crude estimate suggests that the production of 
30keV electrons on sub-second timescales is easily compatible with the use of 
a linear model.             

We have omitted from our model many effects that could in principle influence 
the field decay time. These include ion gyro radius effects, equilibrium 
currents and flows, longitudinal magnetic fields, Hall currents, nonlinear effects, 
pressure gradients, the effects of three dimensional geometry, compressibility 
and kinetic effects. For example, both McClymont \& Craig (1996) and Biskamp 
et al. (1997), using very different physical models, have found that the 
reconnection rate can be strongly affected by the presence of a longitudinal 
magnetic field. Our purpose in this paper has been to achieve a reasonably 
complete understanding of a relatively simple physical model, with the 
minimum number of free parameters. To achieve this understanding, 
we have made full use of the spectral analysis presented in our earlier paper 
(McClements \& Thyagaraja 2004), although the complementary initial value 
approach was essential in order to demonstrate some of our key results, 
notably collisionless continuum damping of the field energy on the Alfv\'en 
timescale. The scheme we have developed could be readily extended to 
incorporate most of the additional effects listed above. The results in this 
paper demonstrate that resistivity and electron inertia alone are sufficient 
to endow perturbed magnetic X-points with a rich physical structure. 

This work was funded by the United Kingdom Engineering
and Physical Sciences Research Council. Helpful conversations with Per
Helander are gratefully acknowledged.

\section*{Appendix A: Relation between Electron Kinetic Energy and Field Energy}

We can establish a useful relation between the field energy ${\cal E}_{\rm f}$ 
and the electron kinetic energy ${\cal E}_{\rm ke}$ as follows. The latter is
given by
$$ {\cal E}_{\rm ke} = {1\over 2}\delta_e^2\int_0^1{1\over r}\left[{\partial
\over\partial r}\left(r\varphi\right)\right]^2dr. 
\eqno ({\rm A}1) $$
where $\varphi = -\partial\psi/\partial r$ is the (purely azimuthal) 
perturbation to the magnetic field. This expression prompts us to consider the 
Rayleigh quotient $R[\varphi]$ defined by the expression 
$$ R[\varphi] = {\int_0^1{1\over r}\left[{\partial\over\partial r}(r\varphi
)\right]^2dr\over \int_0^1\varphi^2rdr}. \eqno ({\rm A}2) $$
Following the ideas underlying Wirtinger's inequality (Hardy, Littlewood, \& 
Polya 1967), we attempt to minimize this functional over all test functions
$\varphi$ that vanish at $r=0$ and $r=1$. According to the Rayleigh-Ritz
principle frequently employed in wave mechanics (e.g. Mathews \& Walker 1964), the 
minimum is attained for $\varphi$ satisfying the Euler-Lagrange equation   
$$ {\partial\over\partial r}\left[{1\over r}{\partial\over\partial r}
\left(r\varphi\right)\right]+\lambda_0^2\varphi = 0, \eqno ({\rm A}3) $$
where $\lambda_0^2 > 0$ is the smallest eigenvalue of this equation.    
Denoting the corresponding eigenfunction by $\varphi_{\rm min}$, it is 
straightforward to establish that 
$$ \varphi_{\rm min} = AJ_1(\lambda_0r), \eqno ({\rm A}4) $$
where $J_1$ is the Bessel function of order one and $A$ is a normalization 
constant that can be chosen to be unity without loss of generality. 
Evidently $\lambda_0$ is equal to the first positive zero of $J_1$, i.e. 
$j_{1,1} \simeq 3.8$ (e.g. Abramowitz \& Stegun 1965).  
From equations (A2) and (A3), integrating the numerator of the former by
parts and using the boundary conditions $\varphi = 0$ at $r=0$ and $r=1$, 
we infer the inequality 
$$ R[\varphi] \ge \lambda_0^2=j_{1,1}^2 \simeq 14.4. \eqno ({\rm A}5) $$
The equals sign in this result applies if and only if $\varphi=
\varphi_{\rm min}=J_1(\lambda_0r)$. Since the field energy is given by
$$ {\cal E}_{\rm f} = {1\over 2}\int_0^1\left({\partial\psi\over\partial r}
\right)^2rdr={1\over 2}\int_0^1\varphi^2rdr, \eqno ({\rm A}6) $$
it follows that $R[\varphi] = {\cal E}_{\rm ke}/({\cal E}_{\rm f}\delta_e^2)$
and hence 
$$ {\cal E}_{\rm ke} \ge \lambda_0^2\delta_e^2{\cal E}_{\rm f}. \eqno 
({\rm A}7) $$  
Thus, for any perturbation to the equilibrium X-point magnetic field, the
electrons must have finite kinetic energy at all times. 

In general, $\varphi$ can be expanded in a Fourier-Bessel series of the form
$$ \varphi = \sum_{n=0}^{\infty} \sqrt{2}a_nJ_1(\lambda_nr)/J_2(\lambda_n), 
\eqno ({\rm A}8) $$
where $\lambda_n$ is the $(n+1)$-th positive zero of $J_1$, the $a_n$ are
constants, and we have used the relation (Whittaker \& Watson 1927) 
$$ \int_0^1J_1(\lambda_nr)J_1(\lambda_mr)rdr = J_2^2(\lambda_n)\delta_{mn}/2, 
\eqno ({\rm A}9) $$
$J_2$ being the Bessel function of order two and $\delta_{mn}$ the Kronecker 
delta symbol, to ensure that the eigenfunctions form a complete orthonormal 
set. The Rayleigh quotient is then given by 
$$ R[\varphi] = {\sum_{n=0}^{\infty}a_n^2\lambda_n^2\over \sum_{n=0}^{\infty}
a_n^2}, \eqno ({\rm A}10) $$
i.e. a weighted mean eigenvalue of the differential operator in equation (A3),
the weights in question being the squares of the Fourier-Bessel amplitudes 
$a_n$ of the perturbed field $\varphi$. Since the $\lambda_n$ form a monotonic 
increasing sequence and the index $n$ counts the number of nodes of the 
eigenfunction, it follows that $R$ can be regarded as a measure of the 
effective radial wave number of $\varphi$ (cf. Thyagaraja 1979). The initial 
value of this measure relates to the localization of the initial state. If
$\varphi$ became more localized or developed more nodes as the system 
evolved, it would be characterized by relatively high eigenvalues
$\lambda_n^2$, and we would then expect a relatively high ratio of electron
kinetic energy to field energy.  
 
\section*{Appendix B: Numerical Algorithms}

To solve equations (15) and (16) numerically it is convenient to introduce new 
variables $u = rv$,  $b = r\partial \psi/\partial r$ and
$$ w =  b - \delta_ { e}^{2} r \frac { \partial }{\partial r }\left(\frac {1 } 
{r}\frac { \partial b} { \partial r}\right). \eqno ({\rm B}1) $$ 
Differentiating equation (15) with respect to $r$ and multiplying by $r$ we 
obtain 
$$ \frac { \partial w} { \partial t} = r \frac{\partial u}{\partial r} +\frac 
{1}{S} r \frac {\partial}{\partial r}\left(\frac { 1} { r} \frac{ \partial b } 
{ \partial r }\right), \eqno ({\rm B}2) $$ 
while the momentum equation becomes
$$ \frac {\partial u} { \partial t} = r  \frac { \partial b} { \partial r }. 
\eqno ({\rm B}3) $$
For the case of finite $\delta_e$ we solve these equations using a staggered 
leap-frog method, in which $u(r,t)$ 
is displaced with respect to $b(r,t)$ and $w(r,t)$ by half the space step, 
$\Delta r$. Labelling the time step by $\Delta t$, and the space and time grid 
points by $i$ and $n$ respectively, we approximate equations (B1) - (B3) using 
the finite difference scheme 
$$ \frac { u_{i+1/2}^{n+1} - u_{i+1/2}^{n}}{\Delta t} = \frac {r_{i+1/2}}
{2\Delta r}\left(b_{i+1}^{n+1}- b_{i}^{n+1}+b^{n}_{i+1}-b_{i}^{n} \right), 
\eqno ({\rm B}4) $$
$$\leftline{$\displaystyle \frac {w_{i}^{n+1}-w_{i}^{n}}{\Delta t} = 
\frac{r_{i}}{2\Delta r}\left(u_{i+1/2}^{n+1}-u_{i-1/2}^{n+1}+u_{i+1/2}^{n}-
u_{i-1/2}^{n} \right)$}$$
$$+\frac {r_i}{S(\Delta r)^{2}}\left[ \frac { b_{i+1}^{n+1}-b_{i}^{n+1}}
{r_{i+1/2}}-\frac{b_{i}^{n+1}-b_{i-1}^{n+1}}{r_{i-1/2}} \right], \eqno 
({\rm B}5) $$
$$ w_{i}^{n+1} = b_{i}^{n+1}- r_{i} \left( \frac{\delta_{e}}{\Delta r} 
\right)^{2} \left[ \frac { b_{i+1}^{n+1}-b_{i}^{n+1}}{r_{i+1/2}}  - 
\frac { b_{i}^{n+1}-b_{i-1}^{n+1}}{r_{i-1/2}} \right].  \eqno ({\rm B}6) $$
Equations (B4) - (B6) are advanced in time using an iterative 
predictor/corrector scheme, with a tri-diagonal matrix algorithm being 
used to solve equation (B6) for $b$ in terms of $w$.

The resistive solutions ($\delta_e=0$) presented in \S\S 3.3 were obtained 
using a code based on a somewhat simpler technique. The discrete 
values of the field variable $b$ are again staggered in space with respect to 
those of the velocity variable $u$, but the scheme is explicit, and thus does 
not require iteration or matrix inversion. As noted in \S\S 3.4, results 
obtained using the two codes are essentially identical whenever $S\delta_e^2 < 
1$. Moreover, for large $S$ and early times, both schemes recover the ideal 
solutions given by equations (26), (27) and (29).        

\section*{References}

\bref{Abramowitz, M., \& Stegun, I. A. 1965 Handbook of Mathematical Functions
(New York: Dover), 390} 

\bref{Birn, J., et al. 2001, J. Geophys. Res. 106, 3715}

\bref{Biskamp, D., Schwarz, E., \& Drake, J. F. 1997, Phys. Plasmas, 4, 1002}

\bref{Bowler, S. 2003, Astronomy \& Geophysics, 44, 6.4}

\bref{Bulanov, S. V., \& Syrovatskii, S. I. 1981, Sov. J. Plasma Phys., 6, 661}

\bref{Case, K. M. 1960, Phys. Fluids, 3, 143}

\bref{Craig, I. J. D., \& McClymont, A. N. 1991, ApJ, 371, L41}

\bref{Craig, I. J. D., \& McClymont, A. N. 1993, ApJ, 405, 207}

\bref{Craig, I. J. D., \& Watson, P. G. 1992, ApJ, 393, 385}   

\bref{Craig, I. J. D., \& Watson, P. G. 2003, Sol. Phys., 214, 131}   

\bref{Hardy, G. H., Littlewood, J. E., \& Polya, G. 1967, Inequalities 
(Cambridge: Cambridge University Press), 184}  

\bref{Heyvaerts, J., \& Priest, E. R. 1983, A\&A, 117, 220} 

\bref{Holman, G. D. 1985, ApJ, 293, 584}

\bref{Kiplinger, A. L., Dennis, B. R., Emslie, A. G., Frost, K. J., \&
Orwig, L. E. 1983, ApJ, 265, L99}

\bref{Lin, R. P., et al. 2002, Sol. Phys., 210, 3} 

\bref{Mathews, J. \& Walker, R. L. 1964, Mathematical Methods of Physics (New York: 
Benjamin), 315} 

\bref{McClements, K. G., \& Thyagaraja, A 2004, Plasma Phys. Control. Fusion, 
46, 39}

\bref{McClymont, A. N., \& Craig, I. J. D. 1996, ApJ, 466, 487}

\bref{McLaughlin, J. A., \& Hood, A. W. 2004, A\&A, submitted}

\bref{Petschek, H. E. 1964, in AAS-NASA Symposium on the Physics of Solar 
Flares (NASA SP-50), 425}

\bref{Porcelli, F., Borgogno, D., Califano, F., Grasso, D., Ottaviani, M., \& 
Pegoraro, F. 2002, Plasma Phys. Control. Fusion, 44, A389} 

\bref{Priest, E., \& Forbes, T. 2000, Magnetic Reconnection: MHD Theory and 
Applications (Cambridge: Cambridge University Press)}

\bref{Ramos, J. J., Porcelli, F., \& Ver\'astegui, R. 2002, Phys. Rev. Lett., 
89, 055002}

\bref{Tandberg-Hanssen, E., \& Emslie, A. G. 1989, The Physics of Solar Flares 
(Cambridge: Cambridge University Press), 175} 

\bref{Thyagaraja, A. 1979, Phys. Fluids, 22, 2093}

\bref{Thyagaraja, A., Loureiro, N., \& Knight, P. J. 2002, J. Plasma Phys., 
68, 363}

\bref{Tsiklauri, D., Nakariakov, V. M., \& Rowlands, G. 2003, A\&A, 400, 1051}

\bref{van Kampen, N. G., \& Felderhof, B. U. 1967, Theoretical Methods in 
Plasma Physics (Amsterdam: North Holland, 153}

\bref{Voitenko, Yu., \& Goosens, M. 2000, A\&A, 357, 1073} 

\bref{Wesson, J. A., 1991, in Plasma Physics and Controlled Nuclear Fusion Research
1990, Vol. 2 (Vienna: International Atomic Energy Agency), 79}

\bref{Whittaker, E. T., \& Watson, G. N. 1927, A Course of Modern Analysis
(Cambridge: Cambridge University Press), 381}

\clearpage

\begin{figure}[htb]
\setlength{\unitlength}{1cm}
\begin{picture}(0.0,18.0)       
\end{picture}
\put(-2.0,0.0){\includegraphics{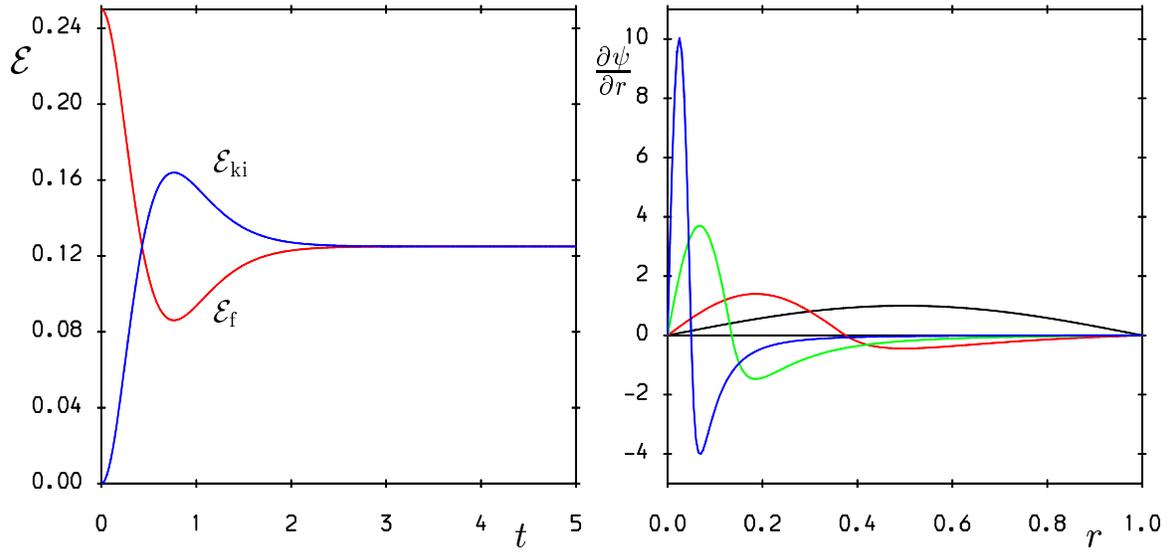}}
\caption{Left plot: time evolution of field energy 
(red curve) and ion kinetic energy (blue curve) in ideal MHD limit for 
$\partial\psi/\partial r\propto\sin(\pi r)$, $v=0$ at $t=0$. Right plot: time 
evolution of $\partial\psi/\partial r$. The curves correspond to $t=0$ 
(black), $t=1$ (red), $t=2$ (green), and $t=3$ (blue).}
\end{figure}

\clearpage

\begin{figure}[htb]
\setlength{\unitlength}{1cm}
\begin{picture}(0.0,18.0)       
\put(-2.0,0.0){\includegraphics{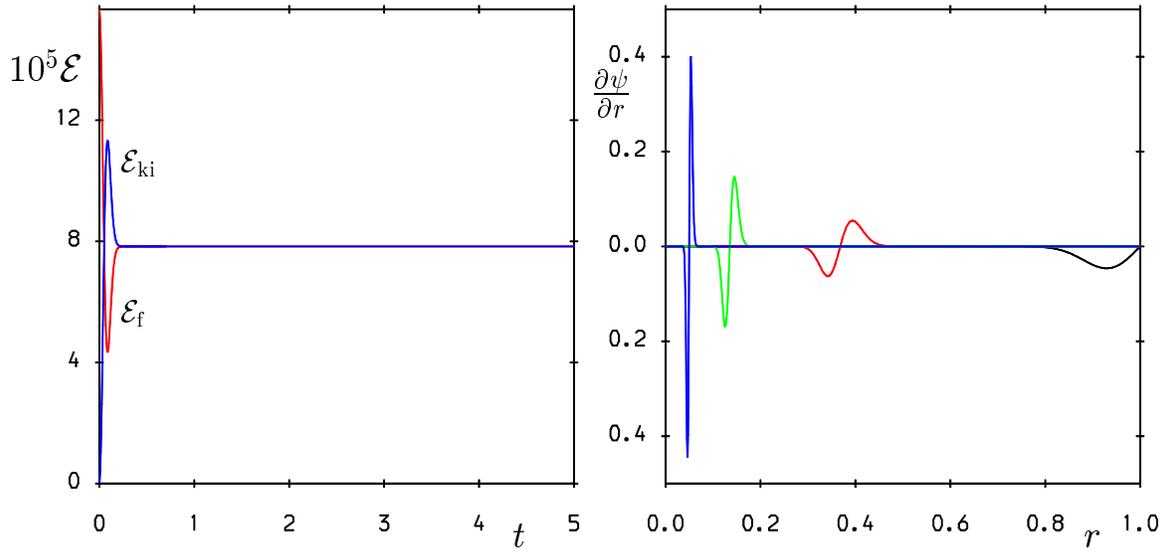}}
\end{picture}
\caption{Left plot: time evolution of field energy 
(red curve) and ion kinetic energy (blue curve) in ideal MHD limit for 
$\psi \propto -\exp[-100(\ln r)^2]$, $v=0$ at $t=0$. Right plot: time 
evolution of $\partial\psi/\partial r$. The curves correspond to $t=0$ 
(black), $t=1$ (red), $t=2$ (green), and $t=3$ (blue).}
\end{figure}

\clearpage

\begin{figure}[htb]
\setlength{\unitlength}{1cm}
\begin{picture}(0.0,19.5)       
\put(-2.0,-6.0){\includegraphics{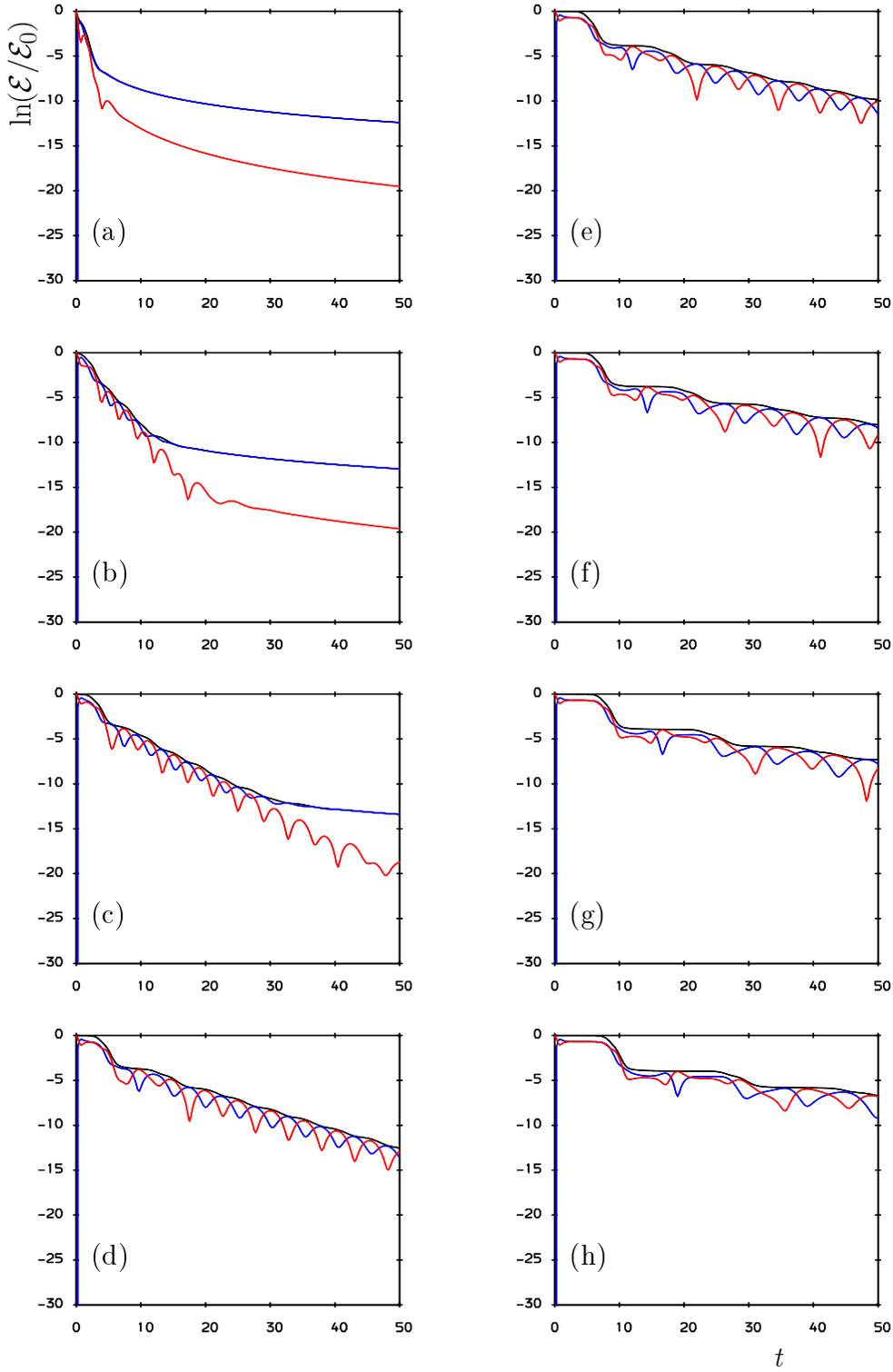}}
\end{picture}
\caption{Time evolution of normalized total energy 
${\cal E}/{\cal E}_0$ (black curves), field energy (red) and kinetic 
energy (blue) for $\delta_e = 0$ and $S=10$ (a),
$10^2$ (b), $10^3$ (c), $10^4$ (d), $10^5$ (e), $10^6$ (f), $10^7$ (g),
$10^8$ (h). The initial conditions are $v=0$ and $\partial\psi/\partial r 
\propto\sin(\pi r)$ }
\end{figure}

\clearpage

\begin{figure}[htb]
\setlength{\unitlength}{1cm}
\begin{picture}(0.0,19.5)       
\put(-2.0,-6.0){\includegraphics{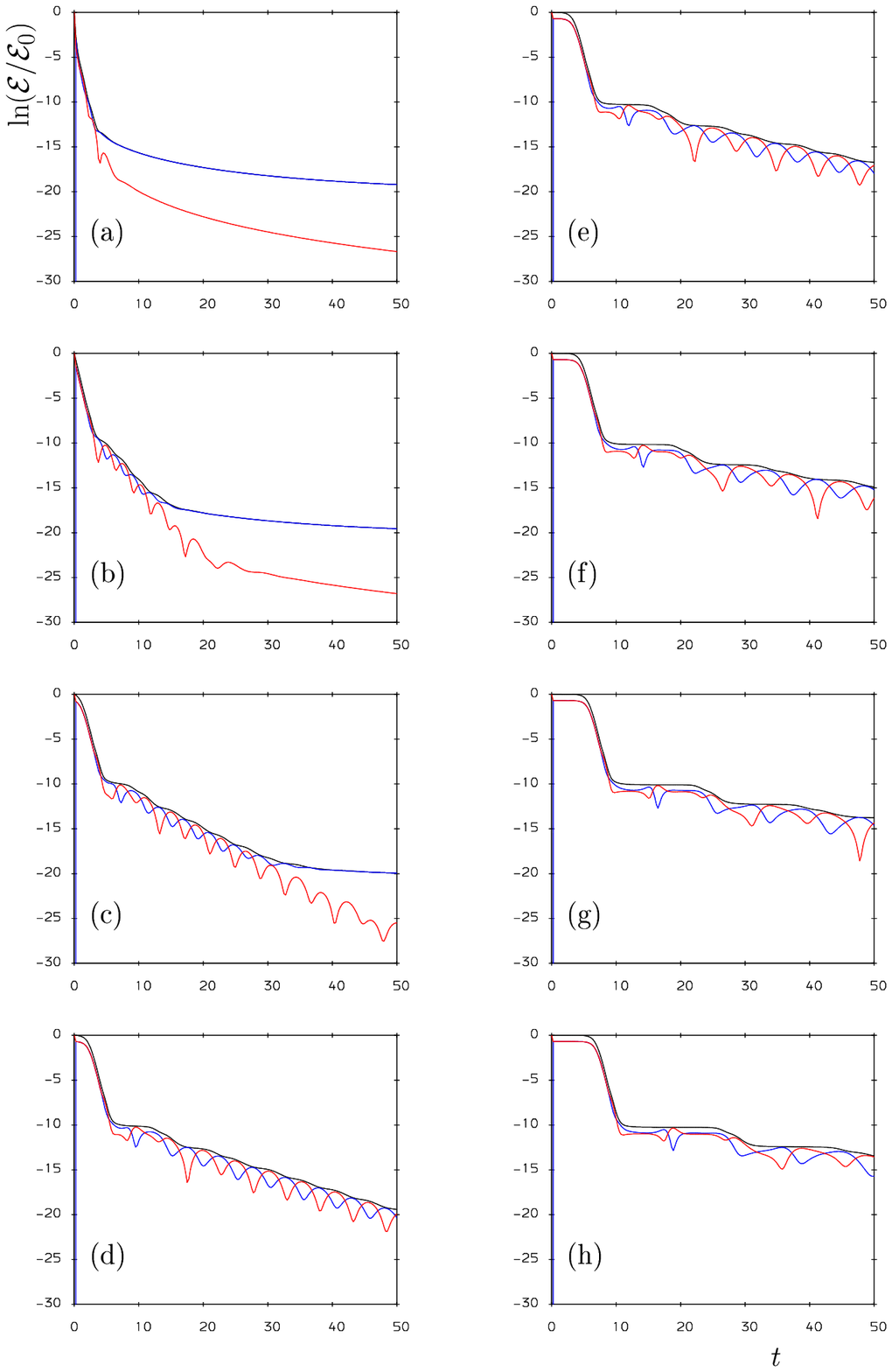}}
\end{picture}
\caption{Time evolution of normalized total energy 
${\cal E}/{\cal E}_0$ (black curves), field energy (red) and kinetic energy 
(blue) for $\delta_e = 0$ and $S=10$ (a),
$10^2$ (b), $10^3$ (c), $10^4$ (d), $10^5$ (e), $10^6$ (f), $10^7$ (g),
$10^8$ (h). The initial conditions are $v=0$ and $\psi \propto -1/
\exp[100(\ln r)^2]$ }
\end{figure}

\clearpage

\begin{figure}[htb]
\setlength{\unitlength}{1cm}
\begin{picture}(0.0,18.0)        
\put(-2.0,-8.0){\includegraphics{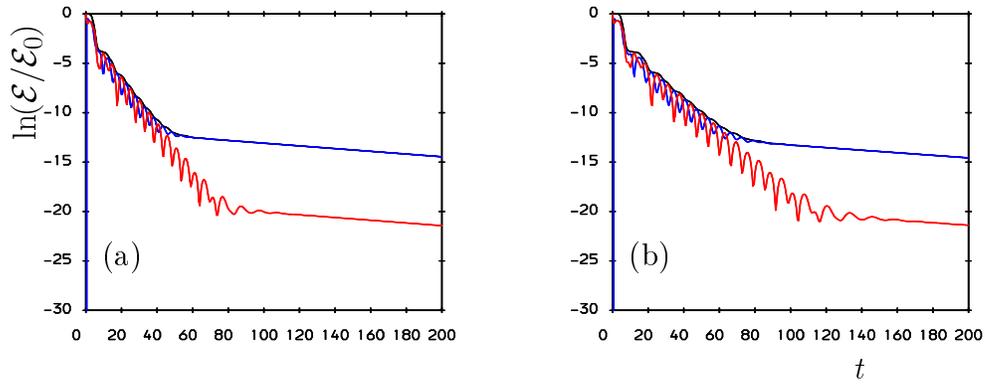}}
\end{picture}
\caption{Time evolution over 200 Alfv\'en periods of 
normalized total energy 
${\cal E}/{\cal E}_0$ (black curves), field energy (red) and kinetic energy 
(blue) for $\delta_e = 0$ and $S=10^4$ (a), $10^5$ (b). The initial conditions 
are $v=0$ and $\partial\psi/\partial r \propto\sin(\pi r)$ }
\end{figure}

\clearpage

\begin{figure}[htb]
\setlength{\unitlength}{1cm}
\begin{picture}(0.0,20.0)       
\put(-2.0,-6.0){\includegraphics{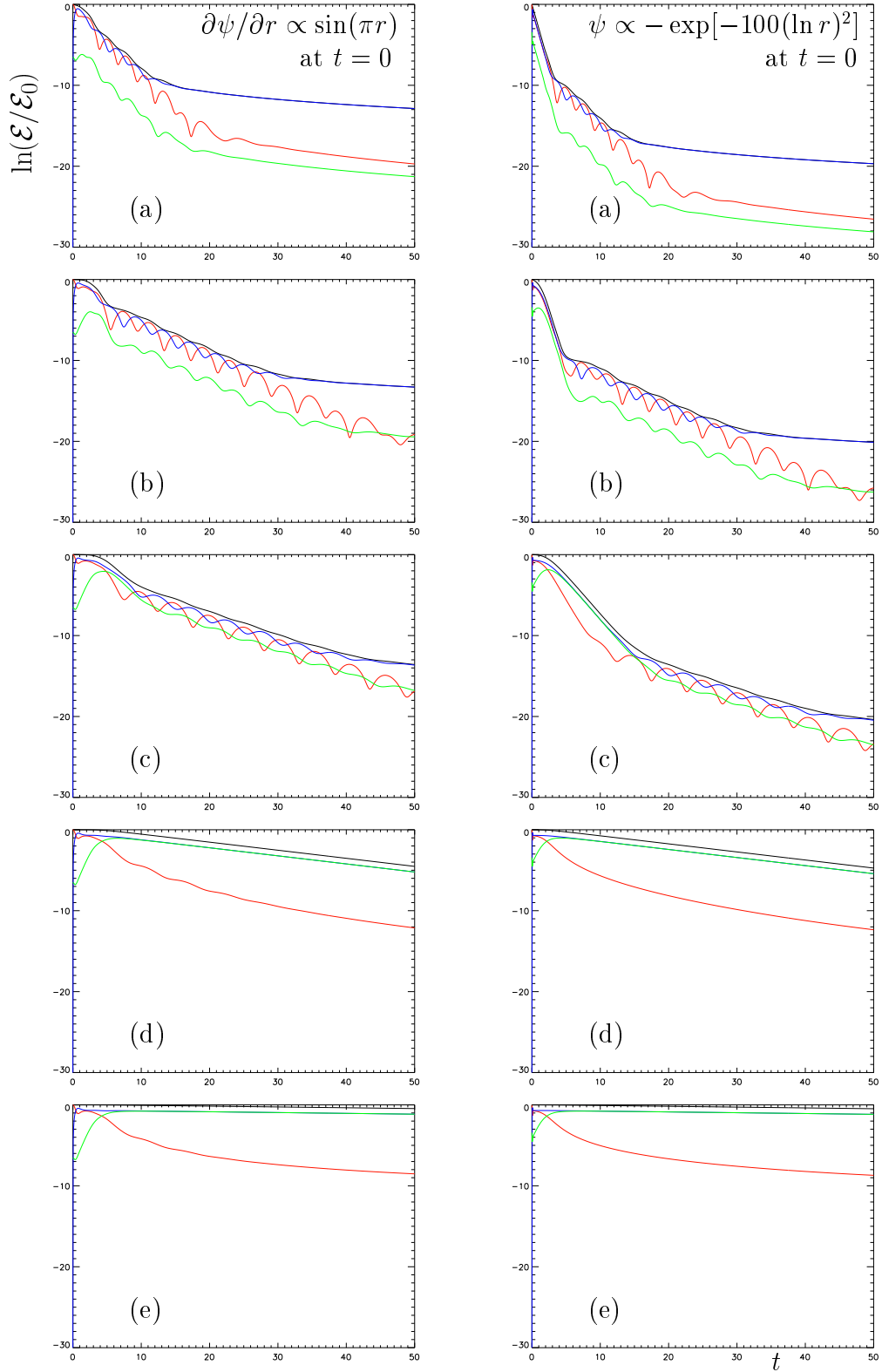}}
\end{picture}
\caption{Time evolution of normalized total energy 
${\cal E}/{\cal E}_0$ (black curves), field energy (red), ion kinetic energy 
(blue) and electron kinetic energy (green) for $\delta_e = 0.01$ and (a) 
$S=10^2$, (b) $10^3$, (c) $10^4$, (d) $10^5$, (e) $10^6$. The initial 
conditions are $v=0$ and $\partial\psi/\partial r \propto\sin(\pi r)$ (left 
plots), $\psi \propto -\exp[-100(\ln r)^2]$ (right plots).}
\end{figure}

\clearpage

\begin{figure}[htb]
\setlength{\unitlength}{1cm}
\begin{picture}(0.0,20.0)       
\put(-1.0,-5.5){\includegraphics{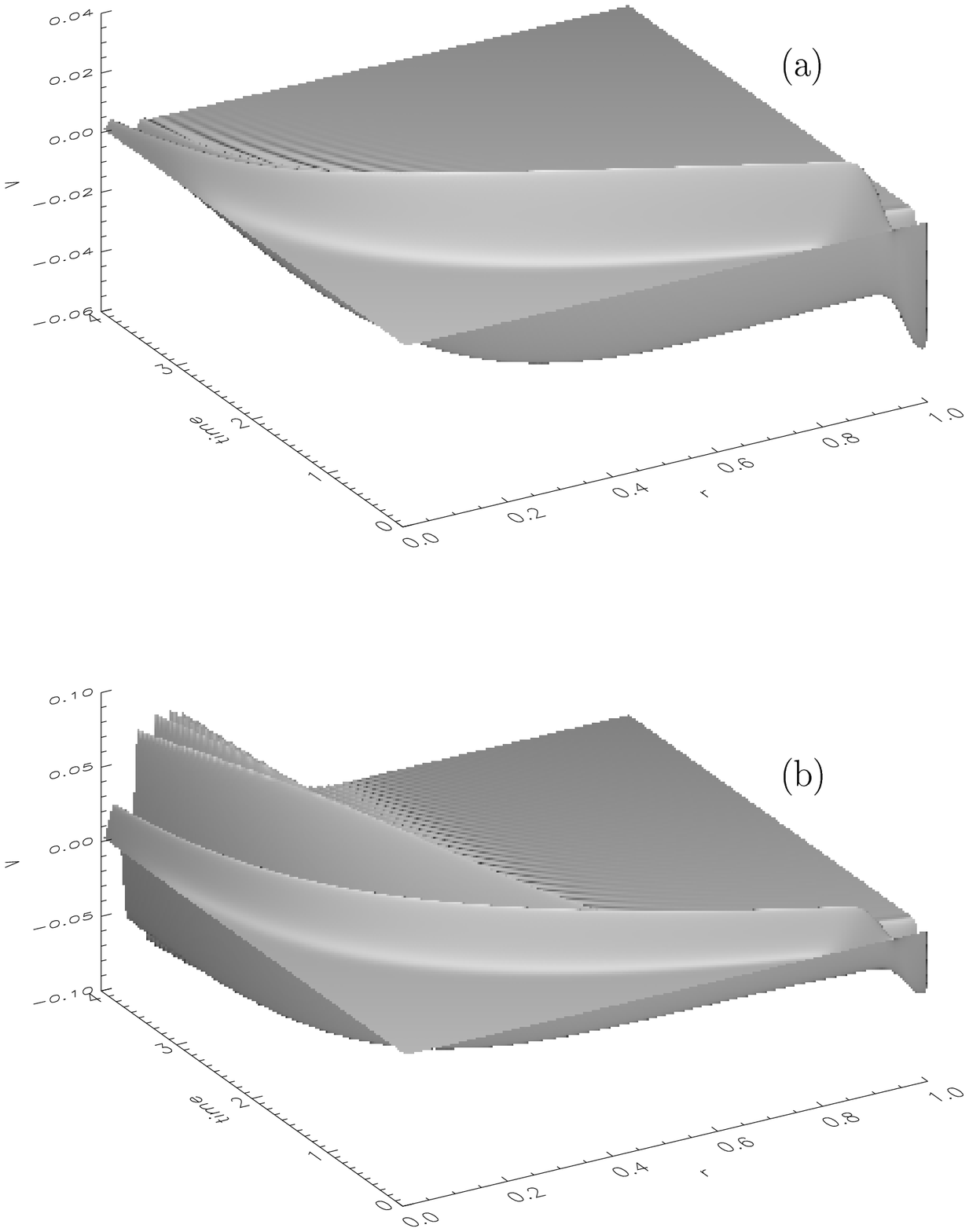}}
\end{picture}
\caption{Space and time evolution up to $t=4$ of velocity 
in simulations with initial conditions $v=0$, $\psi \propto 
-\exp[-100(\ln r)^2]$, $\delta_e=0.01$ and (a) $S=10^3$, (b) $S=10^5$.}
\end{figure}

\clearpage

\begin{figure}[htb]
\setlength{\unitlength}{1cm}
\begin{picture}(0.0,19.0)       
\put(-2.0,-6.0){\includegraphics{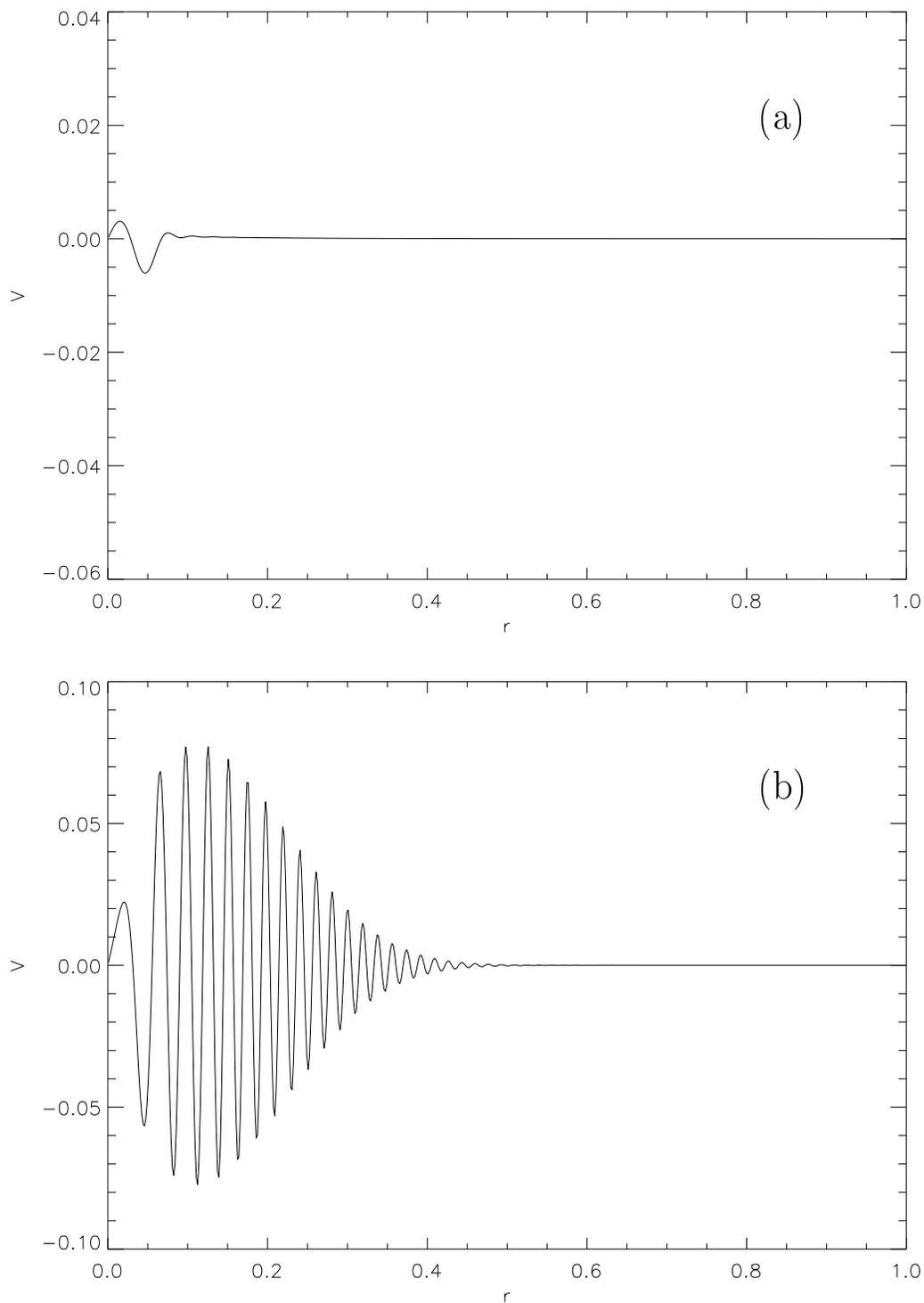}}
\end{picture}
\caption{Velocity profile at $t=4$ in simulations with
$\delta_e = 0.01$ and (a) $S=10^3$, (b) $S=10^5$. The initial conditions are 
$v=0$ and $\psi \propto -\exp[-100(\ln r)^2]$.}
\end{figure}

\end{document}